\documentclass[twocolumn,showpacs,preprintnumbers,amsmath,amssymb,letterpaper]{revtex4}
\usepackage{graphicx}
\usepackage{dcolumn}
\usepackage{bm}
\usepackage{amssymb}
\usepackage{epsfig}

\newcommand{\beq}{\begin{equation}}
\newcommand{\eeq}{\end{equation}}
\newcommand{\bea}{\begin{eqnarray}}
\newcommand{\eea}{\end{eqnarray}}

\newcommand{\fr}[2]{{\frac{#1}{#2}\,}}

\renewcommand{\(}{\left(}
\renewcommand{\)}{\right)}

\begin{document}

\preprint{\hskip5.5in\vbox{BI-TP 2010/14 \\PUPT-2339 \\TUW-10-05}}

\title{Hyperfine Splitting and the Zeeman Effect in Holographic Heavy-Light Mesons}
\author{Christopher~P.~Herzog$^{1}$}
\email{cpherzog@princeton.edu}

\author{Stefan~A.~Stricker$^{2}$}
\email{stricker@hep.itp.tuwien.ac.at}

\author{Aleksi~Vuorinen$^{3}$}
\email{vuorinen@physik.uni-bielefeld.de}

\affiliation{$^1$ Joseph Henry Laboratories, Princeton University, Princeton, NJ 08544, USA }
\affiliation{$^2$ Institut f\"ur Theoretische Physik, Technische Universit\"at Wien, Wiedner Hauptstr.~8-10, A-1040 Vienna, Austria }
\affiliation{$^3$ Faculty of Physics, University of Bielefeld, D-33501 Bielefeld, Germany}

\begin{abstract}
\noindent  We inspect the mass spectrum of heavy-light mesons in deformed ${\mathcal N}=2$ super Yang-Mills theory using the AdS/CFT correspondence. We demonstrate how some of the degeneracies of the supersymmetric meson spectrum can be removed upon breaking the supersymmetry, thus leading to the emergence of hyperfine structure. The explicit SUSY breaking scenarios we consider involve on one hand tilting one of the two fundamental D7 branes inside the internal $\mathbb{R}^6$ space, and on the other hand applying an external magnetic field on the (untilted) branes. The latter scenario leads to the well-known Zeeman effect, which we inspect for both weak and strong magnetic fields.

 \end{abstract}
\date{\today}
\maketitle

\section{Introduction}

A quantitative understanding of the hadron spectrum of QCD is one of the major goals of present day particle physics, and thus any insights that can be obtained from related, analytically more tractable theories are obviously of great value. In an earlier work \cite{Herzog:2008bp}, we investigated the spectrum of heavy-light mesons in strongly coupled, large-$N$ ${\mathcal N}=4$ super Yang-Mills (SYM) theory, to which we added one heavy and one light ${\mathcal N}=2$ hypermultiplet. Investigating the rotational and vibrational modes of open strings stretching between the two D7 branes \cite{Karch:2002sh}, we found a meson spectrum of the form
\beq
M_{hl} = m_h + m_l \, f\left( \frac{J}{\sqrt{\lambda}}, \frac{Q}{\sqrt{\lambda}}, \frac{n}{\sqrt{\lambda}} \right) + {\mathcal O}\left( \frac{m_l^2}{m_h} \right), \label{oldspec}
\eeq
where $m_{h/l}$ stand for the masses of the heavy and light hypermultiplets, $J$ is the angular momentum of the meson, $Q$ an R-charge, and $n$ a quantum number specifying the radial excitation. This result should be contrasted with the predictions of Heavy Quark Effective Theory (HQET) in QCD \cite{Neubert:1993mb}, the heavy quark symmetry of which leads one to expect a heavy-light spectrum of the parametric form
\beq\label{hqet}
M_{hl}=m_h+\Lambda_{\rm QCD}+\frac{\Lambda_{\rm QCD}^2}{m_h}+ {\mathcal O}\left( \frac{\Lambda_{\rm QCD}^3}{m_h^2} \right),
\eeq
where $\Lambda_{\rm QCD}$ is the QCD scale.

The main result of Ref.~\cite{Herzog:2008bp} was to demonstrate the qualitative similarity of the two spectra assuming $m_l$ plays the role of $\Lambda_{\rm QCD}$ in the ${\mathcal N}=2$ theory: In the limit of $m_h\rightarrow \infty$, the excitation energies of the mesons in the SYM theory indeed become independent of the heavy mass and exhibit \textit{fine structure}, represented by the second term of Eq.~(\ref{hqet}). A somewhat surprising difference was, however, seen in the absence of \textit{hyperfine structure} in the ${\mathcal N}=2$ result, \textit{i.e.~}splittings between two mesons that are suppressed by an inverse power of the heavy mass scale. In QCD, this effect is due to the coupling of the heavy quark spin to the light degrees of freedom, while for the SYM theory, its absence can be attributed to the presence of supersymmetry (SUSY), protecting the masses of mesons inside supermultiplets. Similar studies of holographic meson spectra in somewhat different contexts can be found \textit{e.g.~}from Refs.~\cite{Kruczenski:2003be,Erdmenger:2006bg,Jensen:2008yp}. In addition, see Ref.~\cite{Erdmenger:2007cm} for a review.

In the present Letter, we continue our study of holographic heavy-light mesons, but this time concentrate on the effects of SUSY breaking, our goal being to find a simple setup in which to realize the emergence of hyperfine structure. To this end, we investigate two different physical scenarios. First, we study the effects of tilting one of the D7 branes inside the internal $\mathbb{R}^6$ space, so that the branes are no longer parallel but also do not intersect. Second, we investigate a setup in which an external U(1) magnetic field is applied to one or both of the branes, leading to the so-called Zeeman effect (see also Ref.~\cite{Filev:2007gb}). In both cases, we find that the degeneracy of the ${\mathcal N}=2$ SYM case is partially broken, and mass splittings proportional to inverse powers of $m_h$ emerge.

Our paper is organized as follows. In Section II, we present our notation and setup, while Sections III and IV are devoted to studying the tilting of the D7 brane and the Zeeman effect, respectively. In Section V, we finally conclude with a brief discussion of our results.


\section{The Setup}

Here, we briefly review the setup of Ref.~\cite{Herzog:2008bp}, on which we build our current work. From the AdS/CFT correspondence, we know that the field theory we are interested in, ${\mathcal N}=2$ SYM with gauge group SU($N$), is dual to type IIB string theory living in an $AdS_5\times S^5 $ background with the line element
\beq
ds^2 = L^2 \left[ u^2 \eta_{\mu\nu} dx^\mu dx^\nu + \frac{ \delta_{ij} dy^i dy^j}{u^2} \right] .
\label{warpedproduct}
\eeq
Here, the indices $i$ and $j$ run from one to six and $\mu$ and $\nu$ from zero to three, while $L$ is the radius of curvature. The usual radial coordinate of the AdS space, $z$, is related to the $y^i$ through $z^{-2}\equiv u^2 \equiv \sum_i (y^i)^2$.

The dynamics of the D7 branes are governed by the DBI action
\beq
S_{DBI} = -\tau_7 \int d^8\xi \sqrt{-\mbox{det} (G_{ab} + 2 \pi \alpha' {\mathcal F}_{ab}) } , \label{sdbi}
\eeq
where $\tau_7 = 1 / (2 \pi)^7 \alpha'^4 g_s$ is the D7 brane tension, $1/2 \pi \alpha'$ the string tension, $g_s$ the string coupling constant, $G_{ab}$ the induced metric on the D7 brane, and ${\mathcal F}_{ab}$ a field strength corresponding to a gauge field living on the brane. The AdS/CFT dictionary furthermore leads to the relations
\beq
\frac{L^2}{\alpha'} = \sqrt{\lambda} \; \; \; \mbox{and} \; \; \; 4 \pi g_s = g_{\rm YM}^2 ,
\eeq
where $\lambda = g_{\rm YM}^2 N$ is the 't Hooft coupling.

A field theory meson is dual to an open string hanging between two D7 branes, which in our setup are separated by a finite distance proportional to the mass difference of the hypermultiplets. In the ${\mathcal N}=2$ SUSY preserving case, the branes are parallel and we may use the SO(6) isometry group of the $S^5$ to choose the locations of the heavy and light branes as $y^6=0$ and $y^5\equiv y=y_h$ and $y=y_l$, respectively. These parameters are related to the heavy and light quark masses through
\beq
\label{quarkmasses}
m_h = \frac{L^2}{2\pi \alpha'}\; y_h \; ; \;\;\;
m_l = \frac{L^2} {2 \pi \alpha'}\; y_l \, ,
\eeq
while the string equations of motion are obtained from the familiar Nambu-Goto action
\beq
\label{nambugoto}
S_{NG} =
-\frac{1}{2\pi\alpha'} \int d\tau d\sigma \, \sqrt{(\dot X \cdot X')^2 - (\dot X)^2 (X')^2},
\eeq
where $X^A(\tau,\sigma)$ describes the embedding of the string.

The energy spectrum of the mesons consists of the vibrational and rotational modes of the string and, as shown in Ref.~\cite{Herzog:2008bp}, has the form of Eq.~(\ref{oldspec}). As we have not introduced a confinement scale in our setup, $m_l$ takes the place of $\Lambda_{\rm QCD}$ as an infrared cutoff. An important result to recall from Ref.~\cite{Herzog:2008bp} is the degeneracy of the spectrum: Solving for the frequencies of fluctuations in different directions, one finds that the spectra of the $x^i$ ($i=1,2,3$) and $y^6$ fluctuations coincide with each other, but differ from those of the $y^i$ ($i=1,2,3,4$) ones which again are degenerate.

\section{Hyperfine Splitting}

Consider now a slight deformation of the setup introduced in the previous section, in which the embedding equations of the heavy brane are shifted to $y=y_h$ and
\bea
\cos\theta\, y_6 - \sin\theta\, y_4 &=& 0,
\eea
corresponding to tilting the brane in the $y_4,\,y_6$ plane by an angle $\theta$. (From now on, we will write the 4 and 6 as lower indices.) On the field theory side, the redefinition of the $AdS_5\times S^5$ coordinates that would keep the tilted D7 brane at $y_6=0$ corresponds to an SU(4) R symmetry transformation acting on the ${\mathcal N}=4$ scalars and fermions in the \textbf{6} and \textbf{4} representations, respectively. Hence, the Lagrangian of the deformed, non-SUSY theory can be obtained from that of pure ${\mathcal N}=2$ SYM theory by applying suitable transformations on its fields, which has indeed been performed in Ref.~\cite{Pomoni:2010et}.

In Ref.~\cite{Pomoni:2010et}, it was pointed out that for massless and thus overlapping D7 branes, the tilting produces tachyonic modes which translate into a Coleman-Weinberg instability in the effective potential for the fundamental scalars. In our theory, we assume that the D7-branes are separated by a large enough distance that the tachyon is absent.  In field theory terms, this assumption implies that the difference in mass between the hypermultiplets is large compared to the string scale.  However, because SUSY is broken, the force between the D7-branes will not vanish and the form of the potential can be found in Ref.~\cite{Arfaei:1996rg}.
We ignore this force in this section and assume some unspecified physical effect has stabilized the D7-branes at their rotated positions.

On the gravity side, the tilting of the heavy brane affects the spectra of string fluctuations by modifying the boundary conditions that the $y_6$ and $y_4$ fluctuations have to satisfy. The respective Dirichlet and Neumann boundary conditions of these two modes at $y=y_h$ now become
\bea
\cos\theta\, y_6(y_h)-\sin\theta\, y_4(y_h) &=& 0,\\
\sin\theta\, y_6'(y_h)+\cos\theta\, y_4'(y_h) &=& 0,
\eea
which leads to a shift in the frequencies $\omega$ that in the $\theta\rightarrow 0$ limit can be read off from Ref.~\cite{Herzog:2008bp}. To see this shift quantitatively, we now choose the worldsheet coordinate as $\sigma = 1/y$ and assume a time dependence $y_i \sim e^{-i\omega t}$ for the fluctuations. In this case, both fluctuations satisfy the same linearized differential equation
\beq
y_i'' + \frac{2}{\sigma} y_i' = - \omega^2 y_i  , \label{yeq}
\eeq
where $y_i' \equiv \partial_\sigma y_i$ \cite{Herzog:2008bp}.

It is easily verified that the solutions to Eq.~(\ref{yeq}) that satisfy the correct boundary conditions at the untilted light brane can be written in the forms $y_6 = C_6 D(\sigma)$ and $y_4 = C_4 N(\sigma)$, where
\bea
D(\sigma) &=&   \frac{1}{\sigma} \sin(\omega(\sigma-\sigma_l))   ,
\label{myd}
\\
N(\sigma) &=& \frac{1}{\sigma}\left[ \omega \sigma_l \cos(\omega(\sigma-\sigma_l)) + \sin(\omega(\sigma-\sigma_l) )\right]   .
\label{myn}
\eea
We may now write the boundary conditions at the heavy brane as a two-by-two matrix equation for the vector $v\equiv (C_6, C_4)$, $M v=0$, from which we immediately see that the condition for having non-zero solutions is that the determinant of the matrix $M$ vanish,
\bea
\cos^2\theta\, D(\sigma_h) N'(\sigma_h)+\sin^2\theta\, D'(\sigma_h) N(\sigma_h) = 0. \label{eomtilted}
\eea
The solutions to this equation and the corresponding eigenvectors of $M$ correspond to linear combinations of the $y_6$ and $y_4$ fluctuations that are the physical fluctuation modes of the new system.

An observation important for understanding the solutions to Eq.~(\ref{eomtilted}) in the heavy quark limit $\sigma_h\sim 1/m_h \approx 0$ is that because of the prefactor $1/\sigma$ in Eqs.~(\ref{myd}) and (\ref{myn}),
\bea
N'(\sigma_h) &=& -\frac{1}{\sigma_h} N(\sigma_h) + {\mathcal O}(1)  , \\
D'(\sigma_h) &=& - \frac{1}{\sigma_h} D(\sigma_h) + {\mathcal O}(1)  .
\eea
This implies that to leading order in $\sigma_h$, the $\theta$ dependence vanishes from Eq.~(\ref{eomtilted}) and the allowed frequencies are given by the solutions to $N'(\sigma_h)=0$ and $D(\sigma_h)=0$ that can be read off from Ref.~\cite{Herzog:2008bp}. For the $y^6$ case, the unperturbed solutions read
\beq
\omega_n = \frac{\pi n}{\sigma_l - \sigma_h} , \;\; n\in \mathbb{Z}^+ ,
\label{dirnought}
\eeq
while for the $y_4$ fluctuations, we have to solve the transcendental equation
\beq
\frac{\omega \(\sigma_l-\sigma_h\)}{1+\sigma_l \sigma_h \omega^2} = \tan(\omega(\sigma_l-\sigma_h))
\label{neunought}
\eeq
that also leads to a discrete spectrum.

To determine the leading order corrections to the frequency spectra due to the tilting of the heavy brane, we proceed as follows. Anticipating that the eigenvectors of $M$ can at least to leading order still be identified with the $y_6$ and $y_4$ fluctuations, we make a frequency ansatz of the form
\bea
\omega &=& \omega_0 + \omega_1 \fr{\sigma_h}{\sigma_l} + {\mathcal O}\(\fr{\sigma_h^2}{\sigma_l^2}\),
\eea
where $\omega_0$ corresponds to the $\theta=0$ frequencies of Eqs.~(\ref{dirnought}) and (\ref{neunought}). Looking first at fluctuations in the (mostly) $y_6$ direction, we expand Eq.~(\ref{eomtilted}) around $\omega_{0n}\equiv \pi n/(\sigma_l-\sigma_h)$, obtaining
\bea
\omega_n &=&\frac{\pi n}{\sigma_l - \sigma_h} -\frac{n\pi\sigma_h }{\sigma_l^2} \sin^2\theta+ {\mathcal O}\(\fr{\sigma_h^2}{\sigma_l^3}\).
\label{y6fluct}
\eea
For the $y_4$ direction, we similarly get
\bea
\omega_n &=& \omega_{0n}+ \frac{2\sigma_h}{\sigma_l^2} \,{\rm csc}[2\omega_{0n} \sigma_l ]\, \sin^2\theta+ {\mathcal O}\(\fr{\sigma_h^2}{\sigma_l^3}\) ,
\label{y4fluct}
\eea
where the $\omega_{0n}$'s are now obtained from Eq.~(\ref{neunought}).

Eqs.~(\ref{y6fluct}) and (\ref{y4fluct}) describe the energy spectra of our heavy-light mesons.  To make the existence of hyperfine structure clear, recall that $\sigma_h \sim 1/m_h$ and $\sigma_l \sim 1/m_l$.  Recall also that the energy spectra of the heavy-light excitations that correspond to fluctuations in the $x^i$ and $y^i$, $i=1,2,3$, directions were unchanged in the tilting, and are thus given by the $\theta\rightarrow 0$ limit of Eqs.~(\ref{y6fluct}) and (\ref{y4fluct}), respectively. Thus, there exist energy splittings between the $y_4, y_6$ fluctuations and the the $x^i$ and $y^i$ fluctuations of order $m_l^2 / m_h$. It would be a very interesting exercise in perturbation theory to try to produce a similar structure in the energies of the weakly coupled bound states of the deformed theory, proceeding along the lines of Ref.~\cite{Herzog:2009fw}. We, however, leave this investigation for future work.

\section{The Zeeman effect}

Another possibility for breaking SUSY in the setup of Section II is to apply an external magnetic field on one or both of the D7 branes in the system. In general this will lead to a force between the branes, but there are certain configurations, \textit{e.g.~} if the same magnetic field is applied to both branes, where the system will be a stable BPS configuration \cite{Cho:2005aj} . However, in what follows we will use arbitrary magnetic fields and again assume a stabilizing mechanism.

Here, we will for simplicity study a setting in which the U(1) gauge fields living on the branes correspond to a constant magnetic field pointing in the $x^3$ direction, \textit{i.e.}
\beq
2 \pi  F_{(2)} = 2 \pi H \, dx^1 \wedge dx^2  =  \sqrt{\lambda} b \, dx^1 \wedge dx^2,
\eeq
where we have introduced a rescaled field $b \equiv 2 \pi H / \sqrt{\lambda}$. In general, the magnetic field will change the embedding profiles of the branes, which become functions of the radial coordinate of the AdS space, to be determined from the DBI action of Eq.~(\ref{sdbi}) \cite{Filev:2007gb}. The D-brane curvature results in an inequality between the kinetic and bare masses of the quarks, which, however, is suppressed by a factor $H^2/m_{\rm kin}^4$. The suppression implies that unless we consider very large magnetic fields, we may ignore these bending effects at least for the heavy brane.

Starting from the most general case possible, we introduce independent magnetic fields, $H_h$ and $H_l$ (or $b_h$ and $b_l$) for the heavy and light branes, respectively, and study small fluctuations of the string around the unperturbed solution. The unperturbed straight string is orthogonal to the D7 brane at the point where they meet, and thus the fluctuations don't experience the bending of the brane to linear order.  Neglecting bending effects, we may simply repeat the analysis of the previous section. The only difference is that it is now the coupling of the magnetic field to the endpoints of the string (which behave just like charged particles) that changes the boundary conditions and thus the fluctuation spectra.

The new boundary conditions of the string can be read off from the equation
\beq\label{bc}
\partial_\sigma X^\mu+ 2\pi \alpha' {F^\mu}_\nu \partial_\tau X^\nu=0 ,
\eeq
where $X^\mu$ are the coordinates of the string on the world volume of the brane and $F_{\mu\nu}$ is our field strength tensor. Working in the gauge $\sigma=1/y$, we again assume a time dependence of the form $x_i \sim e^{-i\omega t}$, in which case the fluctuations in the $x_1$ and $x_2$ directions satisfy the linearized differential equation
\beq
x_i'' - \frac{2}{\sigma} x_i' = -\omega^2 x_i .
\eeq
This equation has the general solution $x_i = \sum_{j=1}^2 C_{ij} f_j(\sigma)$, where
\beq
f_1 = \sin \omega \sigma - \omega \sigma \cos \omega \sigma \; , \; \; \;
f_2 = \cos \omega \sigma + \omega \sigma \sin \omega \sigma ,
\eeq
using which the boundary conditions of Eq.~(\ref{bc}) can be expressed in the form of a four-by-four matrix equation $Mv=0$, with $v=(C_{11}, C_{12}, C_{21}, C_{22})$.

The condition for the allowed frequencies becomes again that the determinant of the matrix $M$ vanish. While the full expression for the determinant is too messy to reproduce here, we can study various limits thereof. One particularly simple one is that of small $b_l \sigma_l^2$ and $b_h \sigma_h^2$, in which case we easily obtain
\beq
\omega_{n\pm} = \frac{\pi n \pm (b_l \sigma_l^2 - b_h \sigma_h^2)}{\sigma_l - \sigma_h}  , \label{smallb}
\eeq
the $b_l=b_h=0$ limit of which naturally agrees with Eq.~(\ref{dirnought}). Moreover, if we set $b_l=0$, then the magnetic field on the heavy brane leads to meson mass splittings proportional to $1/m_h^2$.

\begin{figure*}[t]
{\centerline{\def\epsfsize#1#2{0.65#1}
   a)  \epsfbox{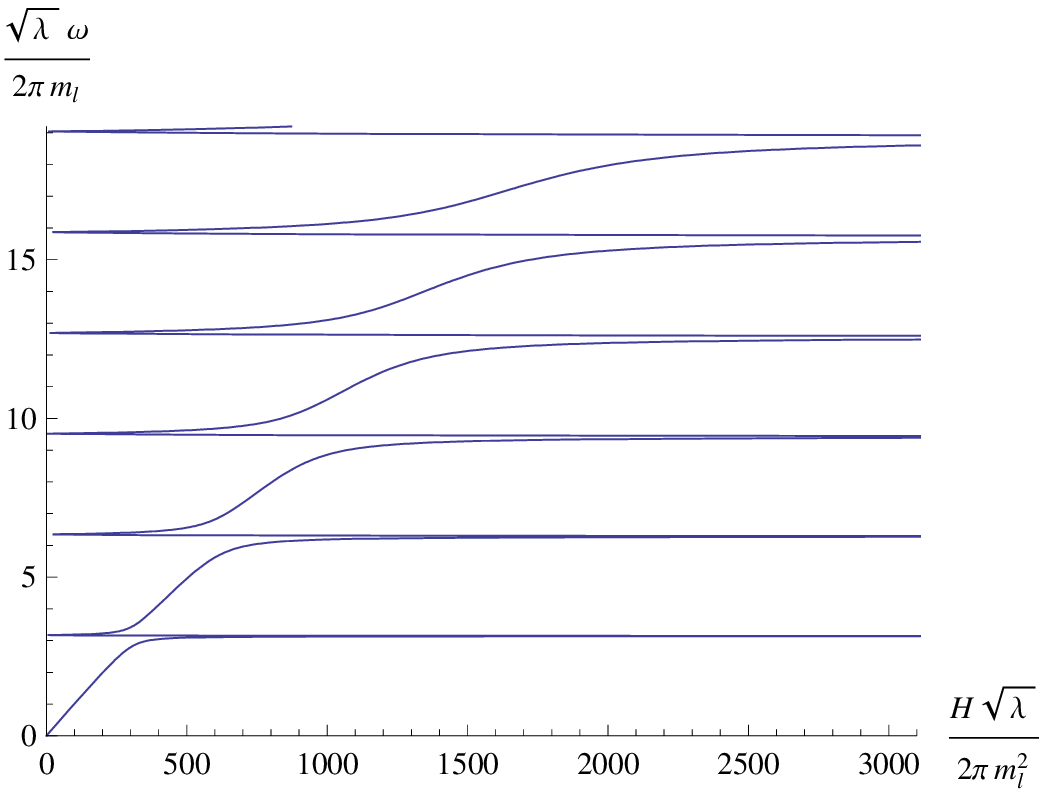}\;
  b)   \epsfbox{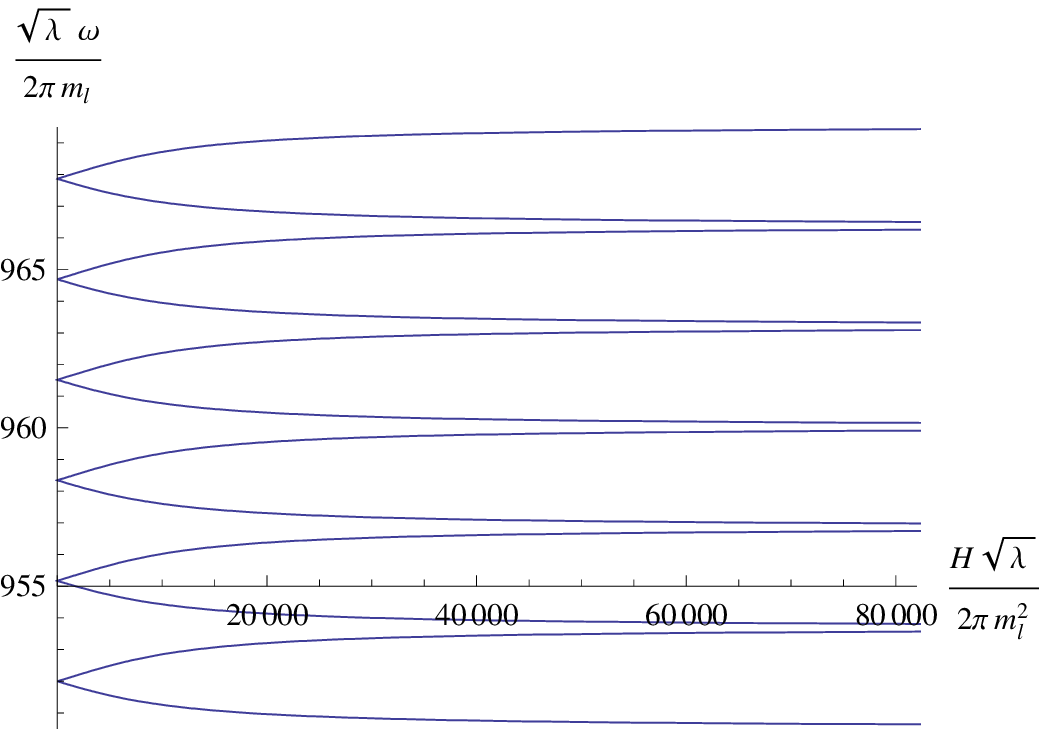}
    }
\caption{The dependence of the $x_1$ and $x_2$ fluctuation frequencies $\omega$ on the magnitude of the magnetic field $H$
 in the case where $m_h/m_l=100$: a) the branches $n=1,2,...,6$ (and half of the zero branch) are displayed; b) the branches $n=301,302,...,306$. In the limit of small $H$, the frequencies can be read off from Eq.~(\ref{smallb}), while for large values of the magnetic field, the behavior of the curves is given by Eq.~(\ref{largeb}).}}
  \label{graph}
\end{figure*}

Finally, we study in some more detail the case where a magnetic field of an arbitrary magnitude is applied on the heavy brane but $b_l=0$. Here, the vanishing of the determinant leads to the condition
\beq
b_h=\pm\,\fr{\omega/\sigma_h}{|1+\omega \sigma_h\cot((\sigma_l-\sigma_h)\omega)|}, \label{bhomega}
\eeq
which we may attempt to invert to find the allowed frequency spectrum. While the small-$b_h$ limit is given by Eq.~(\ref{smallb}), for very large values of the magnetic field we clearly obtain the $b_h$ independent equation
\beq
\tan ((\sigma_l-\sigma_h)\omega) = - \omega \sigma_h , \label{largebeq}
\eeq
which comes with the approximate solutions
\bea
\omega_n &\approx& \frac{\pi n}{\sigma_l} , \;\;n\approx 1, \label{largeb}\\
\omega_n &\approx& \frac{\pi n}{\sigma_l - \sigma_h} - \fr{1}{\sigma_l-\sigma_h}\arctan\bigg[\fr{\pi n}{\sigma_l/\sigma_h-1}\bigg],\;\;n\gg 1. \nonumber
\eea
In Fig.~\ref{graph}, we display the behavior of the frequencies as functions of $b_h$, obtained after numerically inverting Eq.~(\ref{bhomega}). From here, we can identify both the usual Zeeman splitting, described by Eq.~(\ref{smallb}), as well as the subsequent rejoining of the frequencies according to Eq.~(\ref{largeb}).
\vspace{0.8cm}
\section{Conclusions}

In this Letter, we have studied the energy spectrum of strongly coupled heavy-light mesons in a holographic setup. We have observed the emergence of hyperfine structure, \textit{i.e.~}mass splittings suppressed by inverse powers of the heavy mass, upon breaking the SUSY in ${\mathcal N}=2$ SYM theory. Tilting the heavy D7 brane inside the internal $\mathbb{R}^6$ space by an angle $\theta$, we observed splitting terms proportional to $\sin^2\theta/m_h$, while upon introducing a constant U(1) magnetic field on the heavy brane, we found ${\mathcal O}(b/m_h^2)$ effects at weak magnetic fields. Increasing the value of $b$, we finally observed an interesting effect, where the Zeeman split frequencies rejoined in the limit $b\rightarrow \infty$.

While we have chosen to approach the question of hadron spectra in QCD-like theories by considering field theories with well-known gravity duals rather than trying to build phenomenological models for QCD itself, it is interesting to note how some effects characterizing the QCD meson spectrum can be realized in relatively simple settings. The setup involving a tilted heavy D7 brane certainly warrants further work.  In particular, one should try to find a way to stabilize the positions of the two D7-branes.  We are also interested in performing a detailed analysis of the bound states of the theory at weak coupling, building on the results of Refs.~\cite{Herzog:2009fw,Rube:2009yc,Pomoni:2010et}.

\section*{Acknowledgments}

We are indebted to Andreas Karch, Christoph Mayrhofer, Bob McElrath, Radoslav Rashkov, Dam Son, Laurence Yaffe,  and Jure Zupan for valuable discussions and comments. C.P.H.~was supported in part by the National Science Foundation under Grants No.~PHY-0844827 and PHY-0756966, S.A.S.~by the Austrian Science Foundation, FWF, project No.~P19958, and A.V.~by the Humboldt foundation through its Sofja Kovalevskaja program.

\bibliographystyle{h-physrev3}
\bibliography{ref}

\end{document}